\documentstyle[preprint,aps]{revtex}

\begin{document}  
\draft 
  
\title{ Manipulating the scattering length of a
 Bose-Einstein condensate 
 in an amplitude - modulated optical lattice.}  
\author{G.M.Genkin$^*$.}
\address{ Department of Physics and Astronomy, Northwestern University ,
              Evanston , Illinois , 60208 .}

\maketitle 
                 
\begin{abstract}  
               
The 
scattering length
 in a BEC confined
in an amplitude - modulated  
optical lattice
can be manipulated by the modulation
strength.
Two standing laser waves of main and sideband frequencies of an optical lattice
induce a Raman transition;
  due to 
 resonant Raman 
driving the effective
scattering length
  depends on
the modulation strength
 and the detuning from resonance
and can be tuned to a given value.

{*} Electronic address: gena@pluto.phys.nwu.edu.  
\end{abstract}  


   The recent experimental observation of Bose-Einstein condensation
  in a dilute gas of ultracold trapped atoms [ 1-3 ] has generated much interest
  in studying the properties of a Bose-Einstein condensate ( BEC ) and
  manipulating such coherent matter by external electromagnetic fields. 
The experimental observation of Bose-Einstein condensation in a dilute gas of
ultracold trapped atoms [ 1-3 ] has generated much interest in manipulating
such coherent matter by external electromagnetic fields. Recently atoms have
been confined in optical potentials created by standing light waves [ 4, 5 ],
with the wavelength of the optical potential much smaller than the dimensions
of the trap. The unique prospects of this new class of system (the correlated
bosons on a lattice ) became evident.
The possibility of creating optical lattice in trapped Bose - condensed gases
has provided an opportunity to study superfluids in novel situations. The
presence of the lattice leads to a variety of solid - state effects. For
example,
 a transition to a Mott insulator
becomes possible when a superfluid like a BEC is placed in a periodic
potential [4]; the oscillation frequency of the center - of - mass motion
of the condensate is reduced [6] as a result of the enhanced effective mass
of the atoms tunneling between potential wells. By accelerating the optical
lattice [7] Bloch oscillations of the condensate have been observed, and
reducing the amplitude of the lattice optical potential leads to a breakdown
of these oscillations as a result of Landau - Zener tunneling between bands.

  Many physical properties of a dilute gas are determined by the atom - atom
 interaction. In a dilute and cold gas only binary collisions at low energy
 are relevant and these collisions 
are characterized by a single parameter, the $ s $ - wave scattering length.
The confinement of a gas in one or more spatial dimensions can strongly modify
the collisional properties
 (two - body scattering in an optical lattice )
 of atoms [8 - 12].
In Ref.[12] was shown that in an 3D optical lattice a dilute gas is described 
by the
effective atomic scattering length $ a_{eff} $,
modified and tunable by the lattice parameters,
  $$
  a_{eff} =  \frac{ \pi ln 2}{4} a \frac{ d }{ a +
   l_{\ast} } \frac{\hbar}{m \omega l_0^2}, 
   $$
   $$
   l_{\ast} = \frac{ ln 2}{4} l_0 \sqrt{D_0} ,      \eqno (1)
  $$
 where $ a $ is the free - space scattering length, the optical lattice
 potential
  is 
 determined
by a $ V_0 $ and lattice spacing $ d $, and $ l_0 $ is
the size of the ground state wave function in an individual well of an optical
lattice. The wave functions of the relative motion of two particles of mass
$ m $ are characterized by the quasimomentum $ q $, the band index $ s $, and
the energy $ \varepsilon_{s q} = \varepsilon_{s} - t_{s} \cos(q d) $, where
$ t_{s} = \sqrt{D_s} \hbar \omega/\pi, D_s \ll 1 $ is WKB tunneling exponent
between the neighboring wells, $ \omega^2 = V_0/( m d^2 ) , \varepsilon_s $ are
 the energies of states $ s $ in isolated wells.
The effective scattering length $ a_{eff} $ is modified by the optical lattice
if the tunneling between the neighboring lattice sites is sufficiently small and
the new length scale $ l_{\ast} $ contains the tunneling amplitude ( for the
 band index $ s = 0 $ the amplitude $ D_0 $ )
and describes the influence of particle states extending over several
lattice sites on the two - body scattering in an optical lattice.
In Ref. [12] was shown that in the case of attractive interaction, for the 
free - space length
$ a < 0 $, the effective scattering length exhibits resonant behavior and leads
to a geometric resonance at $ l_{\ast} = |a | $, and for $ l_{\ast} > | a | $ the
effective scattering becomes repulsive $ a_{eff} > 0 $.

   The goal of this paper is manipulating of the effective scattering length
   by a time - dependent action of the lattice parameters. 
We consider a manipulation of the scattering length in a BEC in an amplitude
- modulated optical lattice. The such
optical
 lattice can be created by
interfering pair of amplitude - modulated beams.
In general,
there are the two generic types ( a resonant and a nonresonant cases )
of the manipulation.
 The effect of the modulation is to modulate the optical lattice
potential so that it becomes a time - dependent optical lattice potential which
the depth of the optical potential is
  $$
        V_0(t) = V_0(1 + 
        \nu \cos\omega_Mt), \nu < 1,   \eqno (2)
  $$

where $ \nu $  is the modulation strength, $ \omega_M $ is the
modulation frequency. If the period of modulation $ 2\pi/\omega_M $ is
larger the nonequilibrium dynamic time $\tau_{noneq} $, then the time
behavior of a BEC in an amplitude - modulated optical lattice is quasistatic.
The
nonequilibrium dynamic time $ \tau_{noneq} $ for trapped bosonic atoms in an optical
lattice potential was considered in Ref.[11] and was shown that the
characteristic time of the dynamical restoration
 of the phase coherence 
is a Josephson time $ h/J $.

    {\it A resonant case.}

One is a resonant mechanism which operates when the modulation 
 frequency $ \omega_M $ is close to the excited internal state of a BEC.
A resonant mechanism is determined by change in the population
of the ground state due to
a resonant driving.
In a BEC in a periodic potential the characteristic energetic scale of the band
structure [14,15] 
is the recoil energy $ E_R $.
 And, usually, $ E_R $ exceeds $ J $, and the
quasistatic behavior is not valid. However, in this case
is more
convenient to use a picture that an amplitude - modulated field can be
presented as a set of monochromatic fields (main on frequency $ \omega $
 and sideband
on frequencies $ \omega \pm \omega_M $ ). These 
fields can induce a Raman transition between the two
Bloch bands of a BEC $ s = 0 $ and $ s \neq 0 $.
A
resonant field via a two - photon ( Raman ) transition changes the
population of the lowest Bloch band $ s = 0 $ of a BEC. 
In resonant  case the system is excited
by an off - resonant Raman ( two - photon ) driving with the Rabi frequency
$ \Omega_R $. In an amplitude - modulated
optical lattice two standing laser waves of frequencies $ \omega $
 and $ \omega - \omega_M (—\omega_M \ll \omega ) $ 
 drive an atom in a Raman scheme.
  The  effective two -
photon Rabi frequency is
 given in terms of the single - photon Rabi frequencies $ \Omega $ and 
 $\Omega^{\prime} $
 of the fields on the main $ \omega $ and sideband $\omega^{\prime} =
\omega - \omega_M $ 
frequencies as  
  $ \Omega_R = \frac{\Omega \cdot \Omega^{\prime}}{ 2 \Delta} $.
 For the amplitude - modulated laser beams with
the modulation strength $ \nu $ we have for standing waves
$ \Omega_R({\bf x}) = \frac{ \nu \Omega^2({\bf x})}{ 2 \Delta} =
\frac{ \nu V_0({\bf x}) }{ h } $,
here $ \Delta $ is the far - detuning of the laser beams from the atomic 
resonance 
(the closest neighboring optical dipole transition), and
   $ V_0({\bf x}) =\sum_{j=1}^{3} V_{jo} \cos^2 k x_j $ 
 is the optical lattice potential
 with
  wave vectors  $ k = \frac{ 2 \pi }{ \lambda} $ and $ \lambda $  the wavelength of 
   the laser beams    
 on the main frequency, and we are omitted
the terms proportional to the small parameter
$ \Delta k \cdot x = \frac{\omega_M}{ c } x \ll 1 $,
  where $ \Delta k \equiv k^{\prime} - k $, 
 for $ x < \lambda \sim 10^{-4} cm, \omega_M \preceq 10^7 s^{-1} $,
   this parameter is about
$ 10^{-7} $. The eigenstates of a BEC are Bloch waves, and an appropriate
superposition of Bloch states yields a set of Wannier functions which are
localized on the individual lattice sites.
Without a resonant driving the wave functions
in a tight binding model can be represented [12] as
$ w_s ({\bf x})= Z_s \Psi_s ({\bf x}) $, where $ \Psi_s({\bf x}) $ is the wave function of the oscillator 
in a state $ s $, and $ Z_s $ is a normalization factor.
 However, for a resonant driving
 the wave functions are
 $ w_s^{res} ({\bf x}) = \varphi({\bf x}) w_s ({\bf x})$, where
 the function $ \varphi ({\bf x}) $
is the solution
of the Bloch equation of
the standard Rabi problem. 
We consider a steady
state driving regime, in which the function $ \varphi ({\bf x}) $
 is determined on the parameters of the field ( the
Rabi frequency  and the detuning from resonance 
 $ \delta = \omega_M - ( \omega_b - \omega_a ) $ ). 
In order to
   avoid a sufficient population of the excited Bloch band $ s \neq 0 $,
    we are assume
 $ \Omega_R < \delta $, then
$ \varphi^2({\bf x}) \simeq 1 - 
\frac{1}{2} (\frac{\Omega_R({\bf x})}{\delta})^2 $.
   Using the wave functions $ w_s^{res} $, in which
   the size of the ground state oscillator wave function
$ l_0 $ in a lattice is much less than the lattice period
$ d = \lambda/2  $,
   and following Ref. [12] we find the 
   effective scattering
   length  $ a_{ eff}^{res} $
modified by the resonantly amplitude modulation
$$
\frac{ a_{ eff}^{res} - a_{ eff} } { a_{ eff} } \simeq
- (\frac{\nu V_0}{ h \delta})^2.      \eqno (3)
$$

As a result, we can manipulate the 
effective scattering length
in a
BEC in an resonantly amplitude - modulated optical lattice. This tuning 
 is determined by
the modulation 
strength $ \nu $ and the detuning $ \delta $.
 By varying the parameter  $ \nu / \delta $,
 we can tune $ a_{ eff}^{res} $  to a given value.

 {\it A nonresonant case.}
 
 Second is a nonresonant
 mechanism for which we will use
  a direct time - dependent description. Note that such
potential is created by interfering pair of amplitude - modulated laser beams
with the modulation strength $ \nu/2 $
of the electric field amplitude, and was
omitted the quadratic term for the small parameter $ \nu^2/4  $.
A nonresonant low - frequency mechanism of manipulation
 is in operation for a quasistatic regime 
$ \omega_M < 2 \pi \tau_{noneq}^{-1} $, 
therefore, if $ V_0 $ corresponds to a
geometric resonance at $ l_{\ast} = | a | $ for attractive interaction
  then as the time
runs, the system goes from an
attractive interaction
  to a repulsive interaction and back again with
frequency $\omega_M $. Therefore, it is allow to investigate the
nonequilibrium dynamics in a BEC.
In a nonresonant case $ \Omega_R/\delta \rightarrow 1 $
 then $ \varphi({\bf x}) = 1 $
 and we have a 
 time - dependent WKB tunneling exponent $ D_0(t) $.
The WKB tunneling exponent 
is a nonlinear function on  
 the strength of the periodic potential
$ V_0 $ which is proportional to
the laser intensity . If we have a time - dependent lattice laser intensity, then,
although $ \overline {V_0(t)} = V_0 $, due to 
the nonlinearity  of a WKB tunneling exponent
a time - averaged  $ \overline{ D_0(t)} $ will
 be different from a WKB tunneling exponent
$ D_0 $ without a modulation. For a small $ \nu $ we have
 the time - averaged
$ \overline{ D_0( t )} = D_0 + \Delta D_0 $, where
the bar stands for a time average
and
$ \Delta D_0 $
 is proportional to the time - averaged
$ \nu^2 \overline{\cos^2\omega_M t} $ ( the time - averaged of the
modulation harmonic oscillation to the lowest
 even order power ).
For a low - frequency
  nonresonant manipulation we will use
  a
time - dependent potential ( Eq.(2)), and there could be a modulation
frequency $ \omega_M $
 low enough for quasistatic behavior.
  Using the WKB tunneling exponent
  $ D_0 \simeq exp ( -\frac{\pi d}{\hbar} \sqrt{2m V_0} ) $
and using $ V_0(t) $ (Eq.(2)), we have a time - dependent length scale
$ l_{\ast}^{non}(t) $ and, correspondingly, a time - dependent effective
scattering length $ a_{eff}^{non}(t) $.
Calculating the time - average  $ \overline{ D_0( t )} $
 by expanding in a series about the small parameter
 $ 2 m V_0 \pi^2 d^2/ \hbar^2 < 1 $,
we find
 the time - average length scale $ \overline { l_{\ast}^{non}} $, which is
 modified by the nonresonant amplitude modulation
 $$
\frac{ \overline { l_{\ast}^{non} } - l_{\ast} }{ l_{\ast}} \simeq
\frac {\nu^2}{4} \frac {2 m V_0 \pi^2 d^2 }{ \hbar^2}.   \eqno (4)
  $$
If $  \frac {\nu^2}{4} \frac {2 m V_0 \pi^2 d^2 }{ \hbar^2}  $
 is not small, then
 we may expand the expression for the time - dependent $ D_0(t) $ using Bessel
 functions $ I_n( x ) $ and the functions $ \cos^n \omega_M t $.
 In this case instead the parameter $ \frac{\nu^2}{4} 
 \frac {2 m V_0 \pi^2 d^2 }{ \hbar^2}  $ we have
 $ I_2( \nu \frac {\pi d}{\hbar} \sqrt{2m V_0}) $, because,
  usually $ \frac{\pi d}{\hbar} \sqrt{2m V_0} \gg 1 $
  and for $ \nu < 0.4 $ among $ I_n( x < 2 ) $
 the largest is  $ I_2(x) $.

      In summary, the
      effective scattering length
       in a BEC in an
optical lattice can be 
 manipulated 
by an amplitude - modulated manner. 
The effect of the modulation is to modulate the optical lattice
potential so that it becomes a time - dependent optical lattice potential
 depth
  and, correspondingly, the effective scattering length becomes
  time - dependent.
There are two 
 mechanisms of the manipulation.
  One, a resonant mechanism is in operation for the modulation frequency
 closed to the excited internal state of a BEC, and
 two standing laser waves of main and sideband frequencies of an
 amplitude - modulated optical lattice induce a Raman transition.
 Due to resonant Raman driving
   the population of the lowest Bloch band $ s = 0 $ of a BEC changes, 
 therefore, the  effective scattering length
       is also changed. As a result, the effective scattering length
  is other than without a modulation, this
 tuning is
  determined by the square of the modulation strength and the square of the
  detuning.
     Second, a nonresonant low - frequency  mechanism, for which we use
           a direct time - dependent description. Due to
                 the nonlinear dependence
  of the WKB tunneling amplitude on the optical lattice depth
 for the harmonic modulation of the optical lattice
 depth the time - averaged  
 tunneling exponent $ \overline{ D_0( t )} $
  is different from $ D_0 $  without a
 modulation.
  The proposed
mechanism of manipulating the effective scattering length
 makes it
possible to tune it to a given value.
As a result, although a time - averaged optical lattice potential with
a harmonic modulation is equal
to an optical lattice potential without a modulation, 
the effective scattering length
 in a BEC 
confined in an amplitude - modulated optical lattice can be manipulated by
the modulation strength.
In conclusion,
the nonequilibrium dynamics of a BEC can be investigated by manipulating
the effective scattering length of
 a BEC in an amplitude - modulated optical
lattice.

           I thank 
           A.J.Freeman for encouragement.

\end{document}